\documentclass[twocolumn,showpacs,aps,amsmath,amssymb,superscriptaddress]{revtex4}
\usepackage{graphicx,bm,amsmath,amssymb,natbib}

\begin{document}

\title{Detecting the Exchange Phase of Majorana Bound States in a Corbino Geometry Topological Josephson Junction}

\author{Sunghun Park}
\affiliation{Institute for Mathematical Physics, TU Braunschweig, D-38106 Braunschweig, Germany}
\author{Patrik Recher}
\affiliation{Institute for Mathematical Physics, TU Braunschweig, D-38106 Braunschweig, Germany}
\affiliation{Laboratory for Emerging Nanometrology Braunschweig, D-38106 Braunschweig, Germany}

\date{\today}

\begin{abstract}
A phase from an adiabatic exchange of Majorana bound states (MBS) reveals their exotic anyonic nature. 
For detecting this exchange phase, we propose an experimental setup consisting of a Corbino geometry 
Josephson junction on the surface of a topological insulator, in which two MBS at zero energy can be 
created and rotated. We find that if a metallic tip is weakly coupled to a point on the junction, 
the time-averaged differential conductance of the tip-Majorana coupling shows peaks at the tip voltages
$eV = \pm (\alpha - 2\pi l) \hbar/ T_J$, where $\alpha = \pi/2$ is the exchange phase of the two circulating MBS,
$T_J$ is the half rotation time of MBS, and $l$ an integer. This result constitutes a clear experimental signature of Majorana fermion exchange.
\end{abstract}

\maketitle

Majorana fermions are charge-neutral quasiparticles obeying non-Abelian exchange statistics that may be useful 
for quantum computation~\cite{Kitaev2001, Hasan2010, Alicea2012, Beenakker2013, Nayak2008, Moore1991, Ivanov2001}. 
Over the past years, they have been predicted to appear in certain condensed matter systems, like topological insulators~\cite{Hasan2010, Fu2008}, 
semiconductors with spin-orbit interations~\cite{Sau2010, Oreg2010}, and magnetic atom chains~\cite{Choy2011, Nadj-Perge2013, Nadj-Perge2014}, in proximity to {\it s}-wave superconductors, 
and the search for experimental signatures of their unusual features has been intensified. 
For example, the tunneling conductance between a metallic lead and a Majorana fermion shows a zero-bias peak~\cite{Law2009, Flensberg2010}, 
and signatures for this prediction have been identified in experiments~\cite{Mourik2012, Das2012, Deng2012, Lee2014}. 
Other schemes, such as interferometry in a Dirac-Majorana converter~\cite{Fu2009PRL, Akhmerov2009} and a Majorana-mediated Josephson 
effect~\cite{Fu2009PRB, Ioselevich2011, Jiang2011, Sacepe2011, Williams2012, Veldhorst2012, Rokhinson2012}, have also been explored. 
So far, however, most of the signatures that are predicted or observed are attributed to Majorana 
features associated with charge neutrality and fermion parity anomaly, and less attention has been paid 
to find signatures from the exchange statistics, although several schemes to realize 
exchange or braiding operations have been suggested~\cite{Fu2008, Alicea2011, Heck2012, Li2014, Karzig2015}. 

Exchange statistics of Majorana fermions, which differs from that of fermions or bosons, can provide 
alternative routes of detection. An adiabatic exchange of two Majorana fermions $\gamma_1$ and $\gamma_2$ 
leads to the transformation $\gamma_1 \rightarrow \gamma_2$ and $\gamma_2 \rightarrow -\gamma_1$, resulting 
in exotic phase factors~\cite{Ivanov2001}. In Ref.~\cite{Grosfeld2011}, a long circular topological Josephson junction 
is considered where the energy spectrum of the junction is found to depend on the exchange phase and fermion parity. 
Finding further experimental signatures of the exchange phase would be an important task for Majorana fermion detection. 

In this work, we propose a transport experiment where the exchange phase of mobile Majorana bound states (MBS) can be 
identified. The setup consists of a Corbino geometry Josephson junction on the surface of a topological insulator. 
By solving the Bogoliubov-de Gennes equation, we show that two spatially separated MBS at zero 
excitation energy appear in the junction--lacking the hybridization between them--when two flux quanta are introduced in the junction.
Their positions can be moved along the circle by changing the superconducting phase difference across the junction, 
allowing for their exchange. If a metallic tip is weakly coupled locally to the junction, and the two MBS 
are rotating adiabatically, the electron tunneling between the tip and MBS occurs 
every half rotation period. In this weak coupling limit, we find that the time-averaged 
differential conductance shows peaks at the tip voltage $eV = \pm (\alpha - 2 \pi l) \hbar/ T_J$ 
with $\alpha=\pi/2$ the exchange phase~\cite{Mechanism}. This provides direct experimental signatures of the nature of 
Majorana fermion exchange statistics via a dc charge measurement.

\begin{figure}
\centering
\includegraphics[width=0.35\textwidth]{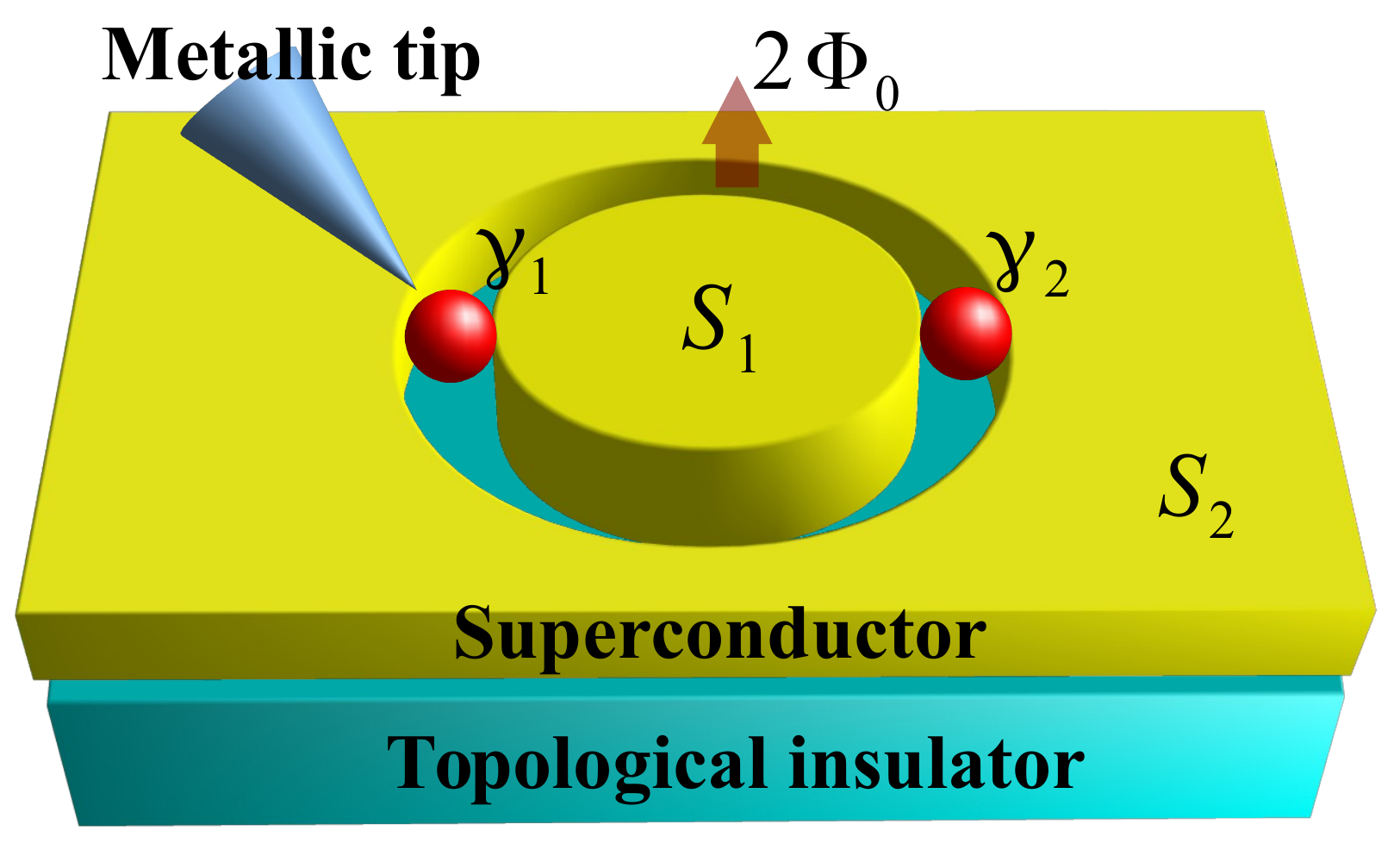}
\caption{Schematic of an experimental setup for performing and detecting Majorana exchange. 
Thin-film superconductors labeled $S_1$ and $S_2$ are deposited on the surface of 
a topological insulator, forming a Corbino-geometry topological Josephson junction.  
In the presence of two flux quanta [$2 \Phi_0$ with $\Phi_0 = h/(2e)$], two zero-energy 
MBS (red balls denoted by $\gamma_1$ and $\gamma_2$) appear at 
opposite sides of the junction. Their positions can be moved slowly while maintaining their relative distance 
by applying a small voltage across the junction, enabling us to perform an adiabatic exchange. 
The resulting exchange phase can be identified in the time-averaged differential conductance of 
a metallic tip weakly tunnel coupled to the junction.
}
\label{Fig1:setup}
\end{figure}

{\it Rotating MBS.---}
We first find MBS in a Corbino geometry Josephson junction 
deposited on the surface ($x$-$y$ plane) of a topological insulator. The setup is illustrated 
in Fig.~\ref{Fig1:setup}. The Josephson junction is formed by thin films of inner ($S_1$) 
and outer ($S_2$) {\it s}-wave superconductors and contains two magnetic flux quanta. For simplicity, we assume 
that the distance between $S_1$ and $S_2$ is zero. We consider the case where the radius 
of $S_1$ denoted by $R$ satisfies $\xi \ll R \ll \Lambda$ so that the vector potential 
of the magnetic field can be neglected, where $\xi$ is the superconducting coherence length, 
and $\Lambda$ is the Pearl length~\cite{Clem2010}; neglecting the vector potential is justified 
provided that the flux through the area where the bound states are localized is smaller than 
the flux quantum~\cite{Akzyanov2014}. We also assume a short Josephson junction, i.e.,   
$2 \pi R \ll \lambda_J$ where $\lambda_J$ is the Josephson penetration depth~\cite{Williams2012,Clem2010}.  
Then the Bogoliubov-de Gennes (BdG) equation with Nambu spinors 
$\Psi = (\Psi_{\uparrow},\Psi_{\downarrow},\Psi^{\dagger}_{\downarrow},-\Psi^{\dagger}_{\uparrow})^{T}$ 
of excitation energy $E$ has the form~\cite{Fu2008, Sau2010, Rakhmanov2011} 
\begin{eqnarray}
\mathcal{H}_{\text{BdG}} \Psi(r, \theta) = E \Psi(r, \theta),\label{BdGEqn} \\ 
\mathcal{H}_{\text{BdG}} = 
\left(
 \begin{array}{ccc}
  H - \mu & \,\, \Delta(r, \theta) \mathbb{I} \\
  \Delta^{*}(r, \theta) \mathbb{I} & \,\, \mu - H
 \end{array}
\right).\label{BdGH}
\end{eqnarray}
Here $H = v_F \vec{\sigma} \cdot \vec{p}$ is the single-particle Hamiltonian describing the surface of 
the topological insulator, $\vec{\sigma}=(\sigma_x, \sigma_y)$ are Pauli spin matrices, 
$\mathbb{I}$ is the $2 \times 2$ unit matrix, and $\mu$ is the chemical potential. The induced gap $\Delta(r, \theta)$ in the presence of the two flux quanta 
can be described by the position-dependent form as
\begin{equation}\label{Gap}
\Delta(r, \theta)= 
\left\{ 
 \begin{array}{cc}
  \Delta_0 e^{i \phi_1} & 0 \leq r < R,\\
  \Delta_0 e^{-i 2 \theta + i \phi_2} &  r > R,
 \end{array}
\right.
\end{equation}
where $\phi_1$ and $\phi_2$ are spatially uniform phases in each region, and 
the polar-angle-dependent term $- 2 \theta$ is due to the flux quanta~\cite{Clem2010}.

\begin{figure}
\centering
\includegraphics[width=0.45\textwidth]{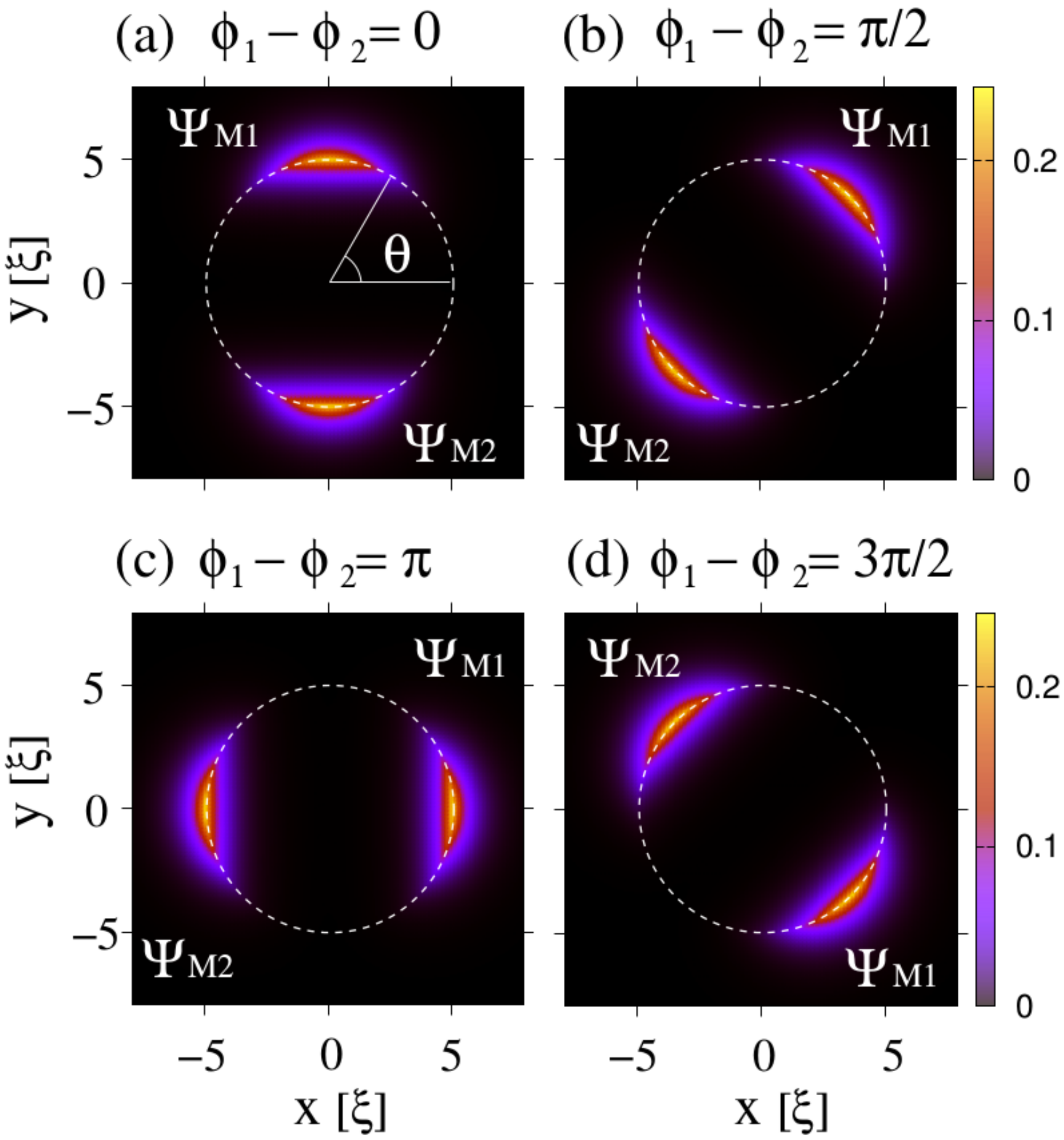}
\caption{(a)-(d) Spatial probability densities of the zero-energy MBS, 
$|\Psi_{M1}|^2$ and $|\Psi_{M2}|^2$, for different values of $\phi_1-\phi_2$ [see Eq.~\eqref{MBSspositions}]. 
The interface of $S_1$ and $S_2$ is depicted by the white dashed circle. 
The MBS are exchanged (or rotated by $\pi$) when $\phi_1 - \phi_2$ varies from $0$ to $2\pi$. 
The parameters are $R=5\xi$ and $\mu=0$, where $\xi = \hbar v_F/\Delta_0$ is the superconducting coherence length.}
\label{Fig2:MBSdensity}
\end{figure}

We solve Eq.~\eqref{BdGEqn} for the case of $\mu =0$ to obtain Majorana zero-energy states, 
$\Psi_{M1}$ and $\Psi_{M2}$, which are spatially separated and orthogonal. They satisfy  $\Xi \Psi_{Mj}=\Psi_{Mj}$, $j=1,2$ 
where $\Xi = \sigma_y \tau_y \mathcal{C}$ is the particle-hole operator and $\mathcal{C}$ the operator for complex conjugation. 
The analytical calculation of the MBS we present in the Supplemental Material~\cite{Supp}. Note that the MBS remain at the zero-energy for 
nonzero $\mu$ (see~\cite{Supp} for the proof). The probability densities of $\Psi_{Mj}$ are shown 
in Fig.~\ref{Fig2:MBSdensity} for different values of $\phi_1 - \phi_2$. 
The result shows that their positions, which are defined by the locations where the probability density takes 
its maximum, are on opposite sides of the junction's circumference, and vary as a function of $\phi_1 - \phi_2$. 
We denote the positions by $\theta = \theta_{\pm}$ with $\theta_{+} (\theta_{-})$ for $\Psi_{M1} (\Psi_{M2})$. 
They are determined by the condition $\Delta \phi= \pm \pi$ and, thus, can be written as 
\begin{equation}\label{MBSspositions}
 \theta_{\pm} = \frac{\phi_2 - \phi_1 \pm \pi}{2},
\end{equation}
where $\Delta \phi \equiv \phi_1 - \phi_2 + 2 \theta$ is the local superconducting phase difference across the junction. 
It is consistent with the previous studies that Majorana fermions can be found in linear~\cite{Fu2008, Potter2013} or 
circular~\cite{Grosfeld2011} Josephson junction when the (local) difference of the superconducting phase is $\pm \pi$. 

The fact that the positions of the MBS are moved by changing $\phi_1 - \phi_2$ 
allows us to perform an adiabatic exchange. When we change adiabatically either $\phi_1$ 
from $0$ to $2 \pi$ or $\phi_2$ from $0$ to $- 2 \pi$, the two MBS rotate by 
$\pi$ in a clockwise direction and are exchanged, as shown in Fig.~\ref{Fig2:MBSdensity}.
During this operation, they continuously evolve as follows~\cite{Ivanov2001}: 
\begin{equation}\label{MBSExchange}
 \Psi_{M1} \rightarrow -s \Psi_{M2}, \,\,\,\,\, \Psi_{M2} \rightarrow s \Psi_{M1},
\end{equation}
where $s=\pm 1$. $s = 1 (-1)$ corresponds to the change of $\phi_1 (\phi_2)$ mentioned above.
We note that this transformation for the exchange is exact as long as the adiabaticity is fulfilled because 
the bound states are exactly at zero energy. 

The adiabatic change of $\phi_1 - \phi_2$ can be achieved experimentally by applying a constant voltage $V_J$ 
across the junction. For the adiabaticity, $e V_J$ should be much smaller than the excitation gap to the excited states. 
Then $\phi_1 - \phi_2$ varies with time $t$ as $\phi_1 - \phi_2 = \phi_0 + 2 e V_J t/\hbar$, 
where $\phi_0$ is constant in time. In this case, $\Psi_{M1}$ and $\Psi_{M2}$ become instantaneous 
zero-energy eigenstates of the time-dependent $\mathcal{H}_{\text{BdG}}[\phi_1(t), \phi_2(t)]$, and their evolution 
during a half rotation time $T_J = \pi \hbar / (e V_J)$ is described as in Eq.~\eqref{MBSExchange}~\cite{periodicity}. 
Equivalently, in terms of Majorana operators, the exchange is expressed as 
\begin{equation}\label{MBSOPExchange}
 \gamma_1(T_J) = -s \gamma_2 (0), \,\,\,\,\, \gamma_2 (T_J) = s \gamma_1(0).
\end{equation}
Here, $\gamma_{j=1,2}(t)= \int d^2r ~\Psi^{\dagger}_{M j}(\vec{r},t) \Phi(\vec{r})$ 
and $\Phi(\vec{r})=(\Phi_{\uparrow},\Phi_{\downarrow},\Phi^{\dagger}_{\downarrow},-\Phi^{\dagger}_{\uparrow})^T$ 
is the four-component field operator. 

{\it Exchange phase and transport.---}
We now consider a normal metallic tip (STM) tunnel coupled to a position of 
the Corbino Josephson junction containing two circulating MBS. 
In this situation, the coupling between the tip and one Majorana state 
becomes significant or suppressed as the Majorana state approaches to or leaves from the tip, respectively, 
while the other Majorana state is on the opposite side of the junction with an exponentially small coupling 
to the tip, leading to occurrence of time-dependent tunneling periodically with period $T_J$.  
In each period, MBS are exchanged. We will see below 
how the resulting exchange phase influences the tunneling differential conductance 
in this setup. 

We perform our calculation in the weak tunneling limit. 
Also, in order to focus on the low-energy regime where the tip-Majorana coupling dominates, 
we assume that the voltage $eV_J (= \pi \hbar / T_J)$ across the junction and tip voltage $eV$ are much 
smaller than the excitation gap. 

Then the low-energy Hamiltonian of our setup is
\begin{equation}
 H(t) = H_N + H_{\text{BdG}}(t) +H_T(t).
\end{equation}
$H_N = \sum_{k,\sigma} \varepsilon_k c^{\dagger}_{k \sigma} c_{k \sigma}$ describes the normal metallic tip 
where $k, \sigma$ represent momentum and spin, respectively, 
and $H_{\text{BdG}}(t) = \frac{1}{2} \int d \vec{r} \Phi^{\dagger}(\vec{r}) \mathcal{H}_{\text{BdG}}(t) \Phi(\vec{r})$. 
The tunneling Hamiltonian between the tip and the junction is
\begin{equation}
 H_T = \sum_{k,\sigma} \int d^2r ~ [t_{k\sigma} (\vec{r}) ~c^{\dagger}_{k \sigma} ~\Phi_{\sigma}(\vec{r}) ~+~ \text{H.c.}].
\end{equation}
As we are interested in low energies, we project $H_T$ onto this low-energy subspace by using 
$\Phi_{\sigma}(\vec{r}) = \sum_{j} ~\Psi_{j,\sigma}(\vec{r},t) \gamma_j(t)$. 
Then we obtain the time-dependent tunneling Hamiltonian,
\begin{equation} 
 H_T (t) = \sum_{j, k,\sigma} [V_{j,k\sigma}(t) ~c^{\dagger}_{k \sigma} ~\gamma_j(t) ~+~ \text{H.c.}],\label{TipJunctionH}
\end{equation}
where $V_{j,k\sigma}(t)$ is the coupling coefficient between the tip and $\gamma_j(t)$ at time $t$, given by
\begin{equation}\label{couplingV}
  V_{j,k\sigma}(t) = \int d^2r ~ t_{k\sigma} (\vec{r})~\Psi_{j, \sigma}(\vec{r},t).
\end{equation}

For further development, we consider a specific model for $H_T(t)$. 
We first define the time sequence $t_q = t_0 + q T_J$ of the coupling with an integer $q$ 
such that if $q$ is even (odd), $V_{1,k\sigma}(t_q)~[V_{2,k\sigma}(t_q)]$ is maximal, 
corresponding to a situation where $\gamma_1(t_q)$ $[\gamma_2(t_q)]$ is located closest to the tip, 
while $\gamma_2(t_q)$ $[\gamma_1(t_q)]$ is on the opposite side of the junction 
with an exponentially small coupling to the tip. Moreover, since MBS are spin polarized, we can 
restrict the consideration to electrons of one spin species of the tip and, thus, effectively deal with spinless electrons. 
By taking nearest-neighbor couplings and writing them 
in terms of $V_{1,k\sigma}(t_0)$ and $\gamma_1(t_0)$ using Eqs.~\eqref{MBSExchange} and \eqref{MBSOPExchange}, 
we obtain the tunneling Hamiltonian at $t=t_q$ as 
\begin{equation}
 H_T(t_q) = H_T(t_0) = \sum_{k} [V_{k} c^{\dagger}_{k} \gamma_1(t_0) + \text{H.c.}], 
\end{equation}
where $V_{k}=V_{1,k\sigma}(t_0)$. 
Next, around $t=t_q$, we assume that the magnitude of the coupling coefficient varies with time in an 
exponential manner, while the phase of the coupling does not change significantly; 
the former assumption is justified by the fact that the MBS show in Fig.~\ref{Fig2:MBSdensity} are exponentially localized 
in the azimuthal direction~\cite{Supp}, and the latter one is a good approximation provided that $\lambda^{-1} \ll T_J$, where 
$\lambda^{-1}$ is the tunneling duration. Then we have 
\begin{equation}
 H_T(t) = \sum_{q,k} e^{-\lambda |t-t_q|} [V_{k} c^{\dagger}_{k} \gamma_1(t_0) +\text{H.c.}].
\end{equation}

The current through the tip is $\langle I(t)\rangle = -e \langle dN_T(t)/dt \rangle$.
$N_T(t) =  \sum_{k} c^{\dagger}_{k}(t) c_{k}(t)$ 
is the tip number operator in the Heisenberg picture with respect to $H(t)$. 
The expectation value $\langle \cdot \rangle$ is taken over a thermal ensemble 
of initial states at time $t=t_0$ being in the far past with $H_T = 0$. We switch on $H_T$ 
at this time~\cite{switchon}. Since tunneling is assumed to be weak, we approximate 
the time evolution operator of $H(t)$ up to first order in $H_T(t)$ in the interaction picture. 
We then obtain the tunneling current 
\begin{align}
 \langle I(t)\rangle =&\frac{1}{i \hbar} \int^{t}_{t_0} dt'  \langle [\hat{I}(t), \hat{H}_T(t')] \rangle \nonumber\\
    =&\frac{2e}{\hbar^2} \text{Re} \Bigg\{ \int^{t}_{t_0} dt' \sum_{k,q,q'} \Gamma_{k,qq'}(t,t') \nonumber\\ 
     &\times [G_{k}(t,t')-\bar{G}_{k}(t,t')] M(t,t') \Bigg\}, \label{LinearresponseI}
\end{align}
in the limit of $\lambda^{-1} \ll T_J$, where the operator notation $\hat{O}$ means an interaction picture operator. 
The current operator is given as
\begin{equation}
 \hat{I}(t) = \frac{e}{\hbar} \sum_{q,k} e^{-\lambda |t-t_q|} [i V_{k} \hat{c}^{\dagger}_{k}(t) \hat{\gamma}_1(t) +\text{H.c.}].
\end{equation}
We note that $\hat{c}^{\dagger}_{k}(t)$ and $\hat{\gamma}_1(t)$ 
evolve in time under $H_N$ and $H_\text{BdG}(t)$, respectively. 
The tunneling magnitude is defined as 
\begin{equation}
 \Gamma_{k,qq'}(t,t') = |V_{k}|^2 e^{-\lambda |t-t_q|} e^{-\lambda |t'-t_{q'}|}. 
\end{equation}
$G_{k}(t,t')$ and $\bar{G}_{k}(t,t')$ are the tip electron Green's functions~\cite{TipGreenftn}, 
and $M(t,t')$ is the Majorana Green's function 
\begin{equation}
 M(t,t') = - i \langle \hat{\gamma}_1(t) \hat{\gamma}_1(t') \rangle.
\end{equation}
This function has the information of the Majorana exchange that is crucial in our proposal. 
Since the tunneling current is exponentially small except for $t=t_q$ and $t'=t_{q'}$ 
due to $\Gamma_{k,qq'}(t,t')$, we can approximate  
$M(t,t') = M(t_q,t_{q'})= -i \langle \hat{\gamma}_1(t_q) \hat{\gamma}_1(t_{q'}) \rangle$.
In the adiabatic limit, the time evolution operator satisfying 
$\hat{\gamma}_1(t+T_J) = U^{\dagger}_M \hat{\gamma}_1(t) U_M$ is~\cite{Ivanov2001} 
\begin{equation}\label{MBSExchangeOP}
 U_M = \text{exp} \left[\frac{\alpha}{2}\gamma_2(t_0) \gamma_1(t_0) \right],
\end{equation}
where $\alpha=s \pi/2$ is the exchange phase where $s=\pm 1$ was defined in Eq.~\eqref{MBSOPExchange}.
From the transformations $\hat{\gamma}_1(t_{q^{(\prime)}}) = [U^{\dagger}_M]^{q^{(\prime)}} \gamma_1(t_0) [U_M]^{q^{(\prime)}}$, 
$M(t_q,t_{q'})$ can be obtained as
\begin{align}
 M(t_q,t_{q'}) =& - \frac{i}{2} \Big\{ [e^{i \alpha (q-q')}+e^{-i \alpha (q-q')}] \nonumber\\
 &+ \mathcal{P}(t_0)[e^{i \alpha (q-q')}-e^{-i \alpha (q-q')}] \Big\} \nonumber\\
 \equiv& - i e^{i \eta \alpha (q-q')}. \label{MBSGreenftn_alpha}
\end{align}
Here, $\mathcal{P}(t_0) = -i \langle \gamma_1(t_0) \gamma_2(t_0)\rangle$ is the fermion parity of 
the junction at $t=t_0$ and is equal to $1 (-1)$ when the two MBS share no (one) fermion, and $\eta=\pm 1$ for $\mathcal{P}(t_0)=\pm 1$.  
By substituting Eq.~\eqref{MBSGreenftn_alpha} into Eq.~\eqref{LinearresponseI}, 
we can obtain the tunneling current $\langle I(t)\rangle$ as a function of time $t$ and the tip voltage $eV$. 

\begin{figure}
\centering
\includegraphics[width=0.45\textwidth]{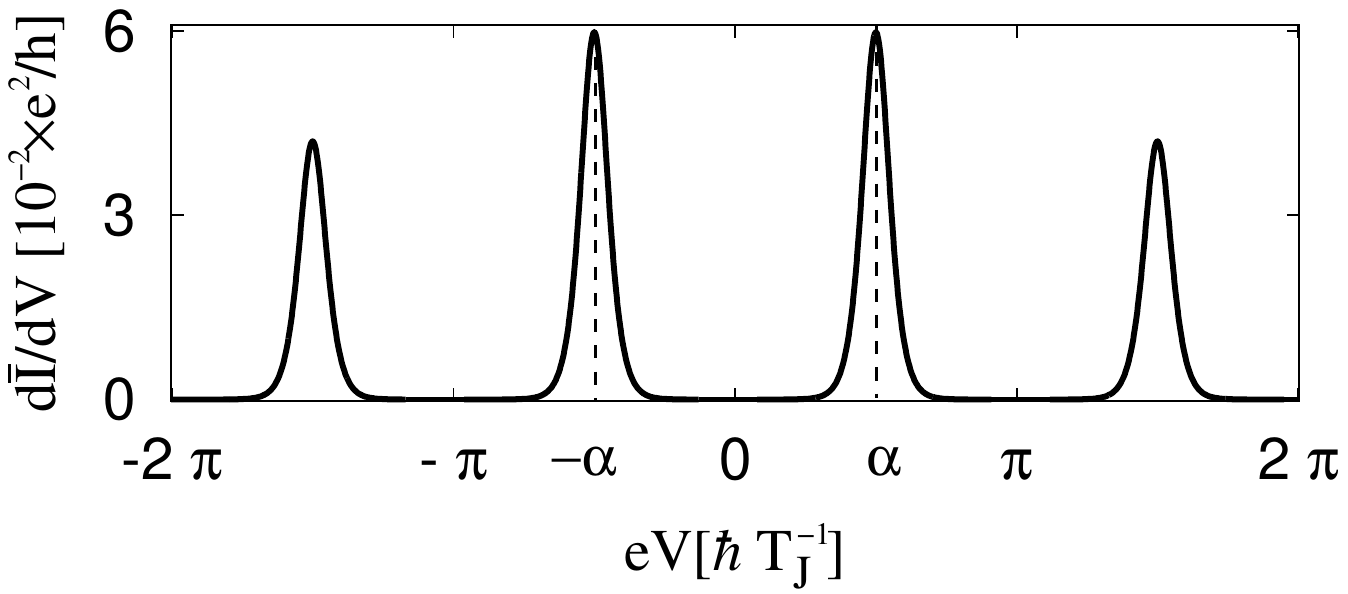}
\caption{Time-averaged differential conductance as a function of $eV$. 
The parameters are $\hbar T_{J}^{-1} = 0.1~\text{meV} = 10^{-1} \hbar \lambda = 10 k_B T = 10 \Gamma$. 
The exchange phase $\alpha =\pi/2$ is used. Around zero voltage, 
the peaks emerge at $eV = \pm \alpha \hbar T_{J}^{-1}$.}
\label{Fig3:dIdV}
\end{figure}

We discuss the voltage dependence of the time-averaged differential conductance $d \bar{I}/dV$~\cite{Supp}. 
It is enough to consider the current for $\alpha = \pi/2$ and $\eta=1$ as it is the same for $\alpha$ and $\eta$ with opposite signs. 
In the limit of $\lambda^{-1} \ll T_J$, we obtain  
\begin{align}
 \frac{d \bar{I}}{dV} =\frac{e^2}{h} \sum_{l} \frac{Z_l(\alpha)}{4 k_B T} \Bigg\{\text{sech}^2 \left[ \frac{V^{-}_l(\alpha)}{2 k_B T} \right] 
 +\text{sech}^2 \left[ \frac{V^{+}_l(\alpha)}{2 k_B T} \right] \Bigg\}, \label{dIdV}
\end{align}
where 
\begin{equation}
 Z_l(\alpha) = 2 \pi \Gamma \left[\frac{2 \lambda T_J}{\lambda^2 T^2_{J} + (\alpha - 2 \pi l)^2}\right]^2,
\end{equation}
and 
\begin{align}
 V^{\pm}_l(\alpha) = \pm eV -\frac{\hbar}{T_J}(\alpha-2\pi l), 
\end{align}
with $l = 0, \pm1, \pm2,...$. Here, $\Gamma = 2 \pi \rho |V_{k}|^2$ is the tunneling energy, and $\rho$ is the tip density of states at the Fermi level. 
We assume the wide-band limit for the tip so that $\Gamma$ becomes energy independent. 
The calculated $d\bar{I}/dV$ is valid for $k_B T \gg |Z_0(\alpha)|$. 
It shows peaks at $eV = \pm (\alpha - 2 \pi l) \hbar/T_J$ with height 
$(e^2/h) Z_l(\alpha)/(4 k_B T)$ as shown in Fig.~\ref{Fig3:dIdV}.

{\it Discussion and experimental feasibility.---}
We have shown that the time-averaged differential conductance between the metallic tip and the Corbino-Josephson junction 
hosting two rotating MBS exhibits peaks at $eV = \pm (\alpha - 2 \pi l)\hbar/T_J$, 
regardless of the fermion parity of the junction. This feature results from the coherent interference in the time-dependent 
tip-Majorana tunneling where the exchange phase $\alpha$ from the half rotation of MBS introduces a relative 
phase between the tunneling at time $t'$ and $t'+T_J$, as shown in Eq.~\eqref{MBSGreenftn_alpha}. We emphasize that, 
different from the case of tunneling between the tip and a static MBS which yields a zero voltage conductance peak~\cite{Flensberg2010}, 
we have conductance peaks at {\it nonzero voltage} for {\it zero-energy} MBS due to the exchange operation. 
The height of the conductance peaks decreases with increasing tip voltage. This is because when the voltage increases, 
the correlation time of electrons in the metallic tip $\hbar/ (eV)$ becomes much smaller than $T_J$, leading to 
the suppression of the interference between tunneling at $t'$ and $t'+T_J$ and, hence, to a decreasing peak height.
We note that the position of the conductance peaks can also be understood by considering an effective Floquet Hamiltonian
as the BdG Hamiltonian is periodic in time with period $T_J$~\cite{Supp}.

We now discuss the experimental feasibility. In our proposal, we require thin-film superconductors whose Pearl length $\Lambda$ 
is much larger than the values of $R$ and $\xi$. Thin-film NbN superconductors have $\Lambda \sim 30$ $\mu$m, $\xi \sim 3.5$ nm, and 
superconducting gap $\Delta_{\text{SC}} \sim 3$ meV~\cite{Lin2013}. Assuming the proximity-induced gap $\Delta_0 \sim 1$ meV, 
the excitation gap of our Corbino-Josephson junction can be estimated by $E_g \sim \sqrt{2 \Delta_0 \hbar \tilde{v}_F/ R} \sim 0.63$ meV 
for $R = 5 \hbar v_F/ \Delta_0$ and $\mu =0$, where $\tilde{v}_F =v_F \Delta^2_0/(\mu^2 + \Delta^2_0)$~\cite{Fu2008}. 
Next, in order to have adiabatic rotation of the MBS and a short tunneling duration, $T_J$ should be long so that 
$\hbar/T_J (= e V_J / \pi) \ll \text{min}\{\hbar \lambda, E_g\}$, but sufficiently short so that $T_J \ll t_{QP}$ 
where $t_{QP}$ is the quasiparticle poisoning time that is in the range of 0.1$-$1 $\mu$s~\cite{Rainis2012}. 
Estimating $\text{min}\{\hbar \lambda, E_g\} \sim 0.63$ meV and $\hbar/t_{QP}\sim 6.6$ neV, gives $\hbar/T_J \sim 10^2$ $\mu$eV 
for adiabatic interference without quasiparticle poisoning. 
In order to observe clearly the conductance peaks shown in Fig.~\ref{Fig3:dIdV}, 
the temperature should be smaller than $\hbar/T_J$ which provides an experimentally feasible temperature range of 10$-$10$^2$ mK.
We finally note that the MBS are robust against weak disorder; i.e., they remain at zero energy~\cite{Supp}, 
and, hence, the disorder has no significant effect on our results in Fig.~\ref{Fig3:dIdV}.

In conclusion, we have proposed an experimental setup where the exchange phase of mobile MBS can be probed 
as a result of their interference effect. Different from the preceding detection schemes 
relying on charge neutrality of zero-energy MBS, our setup allows us to identify the nontrivial MBS statistics, 
providing alternative routes of MBS detection. 

We thank C. W. J. Beenakker, F. Pientka, H.-S. Sim, B. Trauzettel, B. van Heck, and L. Weithofer for valuable discussions. 
P.R. acknowledges financial support from the EU-FP7 Project SE2ND No. 271554 and the DFG Grant No. RE 2978/1-1.

\clearpage

\renewcommand*{\citenumfont}[1]{S#1}
\renewcommand*{\bibnumfmt}[1]{[S#1]}

\widetext
\begin{center}
\textbf{\large Supplemental material for ``Detecting the Exchange Phase of Majorana Bound States in a Corbino Geometry Topological Josephson Junction''} 

\bigskip

Sunghun Park$^1$ and Patrik Recher$^{1,2}$

$^{\it{1}}$\textit{Institute for Mathematical Physics, TU Braunschweig, 38106 Braunschweig, Germany}\\
$^{\it{2}}$\textit{Laboratory for Emerging Nanometrology Braunschweig, D-38106 Braunschweig, Germany}
\end{center}

In this Supplementary Material, we analytically derive Majorana zero-energy states for $\mu =0$ 
and prove that they are still at zero energy for finite $\mu$. 
Next, we provide the details of the calculation of the time-averaged tunneling current $\bar{I}$ in the main text, 
and an alternative explanation for the result shown in Fig. 3 using Floquet theory. 
Finally, we discuss disorder effects. 

\section*{A. Derivation of Majorana zero energy states}\label{sm:MBS}

In this section, we first calculate Majorana zero-energy states, $\Psi_{M1}$ and $\Psi_{M2}$, for $\mu=0$ by solving 
the BdG equation for the Corbino geometry. From Eqs.~(1) and (2) in the main text, the BdG equation for $\mu = 0$ is given by
\begin{equation}
 \left(
 \begin{array}{ccc}
  v_{\text{F}} \vec{\sigma} \cdot \vec{p} & \,\, \Delta(r, \theta) \mathbb{I} \\
  \Delta^{*}(r, \theta) \mathbb{I} & \,\, -v_{\text{F}} \vec{\sigma} \cdot \vec{p}
 \end{array}
\right) \Psi(r, \theta) = E \Psi(r, \theta). \\ 
\end{equation}
This decouples into two sets of equations, one for spin up and one for spin down. 
We consider only the equations with spin down as those for spin up give diverging solutions which are not physical.
Due to the absence of rotation symmetry of the system, the wave function in each region should be given in the 
form of a superposition of different angular momentum eigenstates.
Hereafter we will measure energy and length in units of gap magnitude $\Delta_0$ and 
superconducting coherence length $\xi = \hbar v_{\text{F}}/\Delta_0$, respectively.
With these units, the wave function for the region of $0 \leq r < R$ is expressed as 
\begin{equation}\label{Waveftn1}
 \Psi^{<}(r, \theta) = \sum^{\infty}_{m=-\infty} a_m \Psi^{<}_m(r, \theta). 
\end{equation}
Here, $a_m$ are coefficients to be determined below, and $\Psi^{<}_m(r, \theta)$ is defined as  
\begin{equation}
 \Psi^{<}_m(r, \theta) = \left(
 \begin{array}{cccc}
 0 \\
  e^{i (m+1) \theta} e^{i \phi_1 /2} J_{m+1} (i r) \\
  -e^{i m \theta} e^{-i \phi_1 /2} J_m (i r) \\
 0 
 \end{array}
 \right),
\end{equation}
where $J_m (ir)$ is the Bessel function of the first kind that is regular at the origin $r=0$. 
For $r > R$, the wave function is found as 
\begin{equation}\label{Waveftn2}
 \Psi^{>}(r, \theta) = \sum^{\infty}_{n=-\infty} b_n \Psi^{>}_n(r, \theta),
\end{equation}
with coefficients $b_n$, and 
\begin{equation}
 \Psi^{>}_n(r, \theta) = \left(
 \begin{array}{cccc}
 0\\
  e^{i n \theta} e^{i \phi_2 /2} r H^{(1)}_{n+1} (i r) \\
  -e^{i (n+1) \theta} e^{-i \phi_2 /2} r H^{(1)}_n (i r) \\
  0
 \end{array}
 \right),
\end{equation}
with $H^{(1)}_n(ir)$ the Hankel function of the first kind that goes to zero as $r \rightarrow \infty$.  
In order to obtain solutions, we determine the coefficients $a_m$ and $b_n$ by matching the wave functions $\Psi^{<}(r, \theta)$ and $\Psi^{>}(r, \theta)$ at $r=R$, leading to 
\begin{eqnarray}
 a_l e^{i \phi_1/2} J_{l+1}(i R) = b_{l+1} e^{i \phi_2/2} R H_{l+2}(i R), \nonumber\\
 a_{l+1} e^{-i \phi_1/2} J_{l+1}(i R) = b_l e^{-i \phi_2/2} R H_l(i R), \label{Recurrence1}
\end{eqnarray}
and the following recurrence relations can be derived,
\begin{eqnarray}
 a_{l+2} = e^{i (\phi_1-\phi_2)} \frac{H_{l+1}(i R) J_{l+1}(i R)}{H_{l+2}(i R) J_{l+2}(i R)} a_l, \nonumber\\
 b_{l'+2} = e^{i (\phi_1-\phi_2)} \frac{H_{l'}(i R) J_{l'+2}(i R)}{H_{l'+3}(i R) J_{l'+1}(i R)} b_{l'}, \label{Recurrence2}
\end{eqnarray}
where $l$ and $l'$ range over all integers. 
Eqs.~\eqref{Recurrence1} and \eqref{Recurrence2} allow us to express the coefficients 
$a_{l=\text{even}}$ and $b_{l'=\text{odd}}$ ($a_{l=\text{odd}}$ and $b_{l'=\text{even}}$) in terms of $a_0$ ($a_{-1}$). 
This fact indicates that we can construct two independent solutions, denoted by $\Psi_1$ and $\Psi_2$,
\begin{eqnarray}
\Psi_1 =  \Theta(R-r) \sum_{m=\text{even}} a_m \Psi^{<}_m + \Theta(r-R) \sum_{n=\text{odd}} b_n \Psi^{>}_n, \nonumber\\
\Psi_2 =  \Theta(R-r) \sum_{m=\text{odd}} a_m \Psi^{<}_m + \Theta(r-R) \sum_{n=\text{even}} b_n \Psi^{>}_n, 
\end{eqnarray}
with one unknown coefficient $a_0$ ($a_{-1}$) for $\Psi_1$ ($\Psi_2$), where $\Theta(x)$ is the unit step function. 
According to the particle-hole symmetry, if $\Psi_1$ is a solution, then $\Xi \Psi_1$ is also a solution, 
where $\Xi = \sigma_y \tau_y \mathcal{C}$ is the particle-hole operator with $\mathcal{C}$ the operator of complex conjugation.
This imposes a constraint $\Psi_2 = \Xi \Psi_1$, and hence $a_{-1} = - a^{*}_0$. We choose $a_0 = e^{i(\phi_1-\phi_2)/4}/\sqrt{N}$ 
and $a_{-1} = -e^{-i(\phi_1-\phi_2)/4}/\sqrt{N}$ where $N$ is the normalization constant. 
The following combinations of $\Psi_1$ and $\Psi_2$ give two MBS, 
\begin{equation}
\Psi_{M1} = \frac{1}{\sqrt{2}} (\Psi_1 + \Psi_2), ~~
\Psi_{M2} = \frac{i}{\sqrt{2}} (\Psi_1 - \Psi_2), \label{MBSs}
\end{equation}
which are plotted in Fig. 2 in the main text. 

We show that $\Psi_{M1}$ and $\Psi_{M2}$, centered at $\theta=\theta_{\pm}$, respectively, are exponentially localized 
in the azimuthal direction to justify the assumption in Eq.~(12) in the main text that the coupling between 
the tip and MBS is exponentially small except for $t=t_q$ where $t_q = t_0 + q T_J$ with an integer $q$.
We first solve an effective low-energy Hamiltonian which is the $k \cdot p$ Hamiltonian obtained by Fu and Kane in Ref.~\cite{Fu2008_SM} incorporating 
the spatial variation of a superconducting phase difference due to two flux quanta in the Josephson junction, 
\begin{align}
  \mathcal{H}_{\text{eff}} = 
 \left(
 \begin{array}{cc}
  - i \frac{\hbar \tilde{v}}{R} \partial_{\theta} & \,\, -i \Delta_0 \text{cos}\left(\frac{\Delta\phi}{2} \right) \\
  i \Delta_0 \text{cos}\left(\frac{\Delta\phi}{2} \right) & \,\,  i \frac{\hbar \tilde{v}}{R} \partial_{\theta}
 \end{array}
\right),
\end{align}
where $\tilde{v} = v_{\text{F}} [\text{cos} (\mu W/\hbar v_{\text{F}}) + (\Delta_0/\mu) \text{sin} (\mu W/\hbar v_{\text{F}})] \Delta^2_0/(\mu^2 + \Delta^2_0)$, $\mu$ is the chemical potential, 
$W$ is the distance between two superconductors, and $\Delta\phi = \phi_1 - \phi_2 + 2 \theta$ is the phase difference of the Josephson junction.
The effective Hamiltonian, whose bases are two branches of counter-propagating helical Majorana states after integrating out the transverse (radial) degree of freedom, 
is a good approximation in the limit of $R \gg \xi'$ where $\xi' = \hbar \tilde{v} / \Delta_0$. To calculate approximate zero energy solutions localized at $\theta = \theta_{p}$ 
with $p\in \{+,-\}$, we linearize $\text{cos} (\Delta \phi/2)$ around $\Delta \phi = p \pi$, 
\begin{align}
 \text{cos}\left(\frac{\Delta\phi}{2}\right) &= \text{cos}\left(\frac{p \pi + 2 \delta \theta_p}{2}\right)  \nonumber\\
 &= -p ~\text{sin} (\delta \theta_p) \nonumber\\
 &\approx -p ~\delta \theta_p 
\end{align}
for small $\delta \theta_p$ where $\delta \theta_p \equiv \theta - \theta_{p}$.
Then we obtain the exactly solvable Hamiltonian
\begin{align}
  \mathcal{H}_{\text{eff}} = 
 \left(
 \begin{array}{cc}
  - i \frac{\hbar \tilde{v}}{R} \partial_{\theta} & \,\, -i p ~\Delta_0 \delta \theta_p \\
  i p ~\Delta_0 \delta \theta_p & \,\,  i \frac{\hbar \tilde{v}}{R} \partial_{\theta}
 \end{array}
\right),
\end{align}
thus allowing us to calculate the zero energy solutions or approximate Majorana states which are exponentially localized in the azimuthal direction, given by  
\begin{align}
 \Psi^{\text{approx}}_p (\theta) = \left(\frac{R}{4 \pi \xi'} \right)^{1/4} \text{exp}\left[-\frac{R}{2 \xi'} (\theta - \theta_{p})^2\right]
 \left(
 \begin{array}{cc}
  1 \\
  -p
 \end{array}
 \right),
\end{align}
and probability densities
\begin{align}
 |\Psi^{\text{approx}}_p (\theta)|^2 = \sqrt{\frac{R}{\pi \xi'}} \text{exp}\left[-\frac{R}{\xi'} (\theta - \theta_{p})^2\right]. \label{ApproxMBS}
\end{align}
To compare these approximate solutions with the exact MBS given in Eq.~\eqref{MBSs}, we integrate out the radial degrees of freedom 
of $\Psi_{M1}(r,\theta)$ and $\Psi_{M2}(r,\theta)$, 
\begin{align}
 |\Psi^{\text{exact}}_{M1 (M2)}(\theta)|^2 = \int dr r |\Psi_{M1 (M2)}(r,\theta)|^2. \label{ExactMBS}
\end{align}
\begin{figure}
\centering
\includegraphics[width=0.6\textwidth]{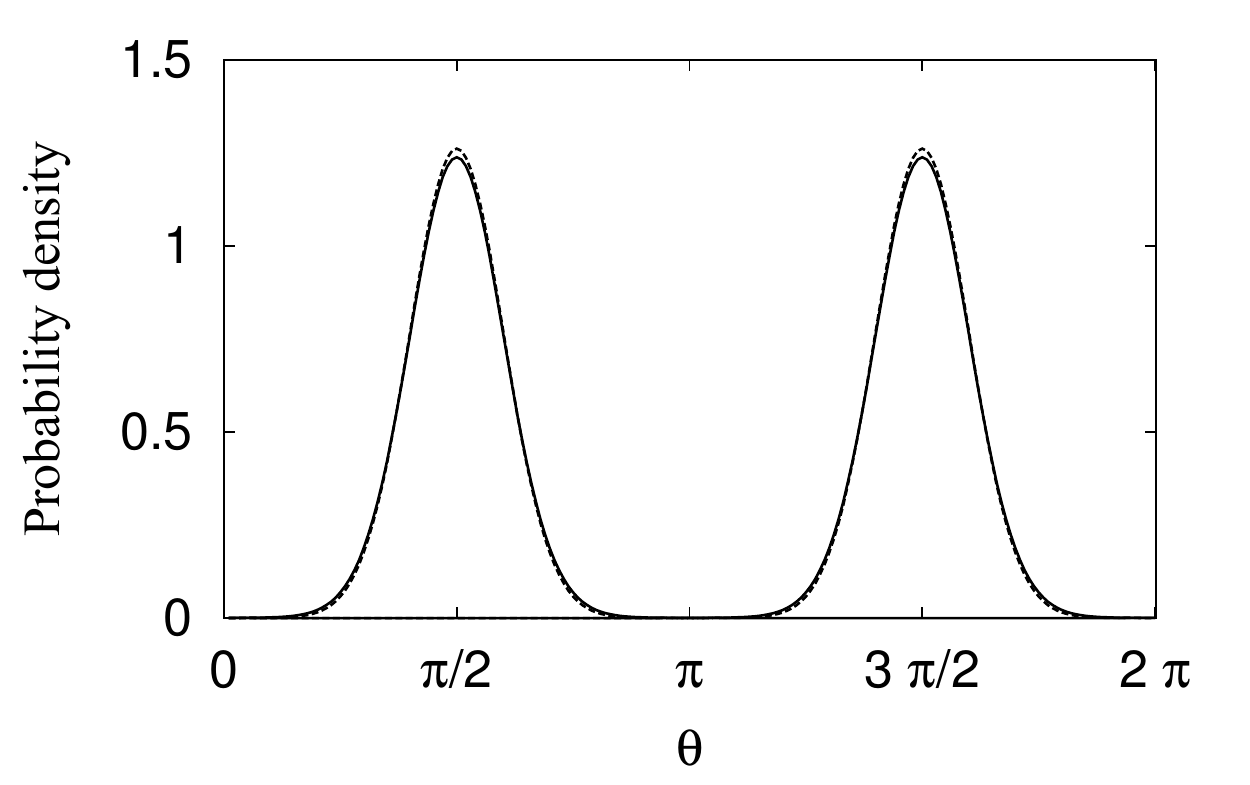}
\caption{Probability densities of MBS as a function of azimuthal angle $\theta$ for the values of $\theta_{\pm} = \pm \pi/2$, $R = 5 \xi'$ and $\mu = W = 0$. 
Probability densities of $\Psi^{\text{exact}}_{j=M1,M2}(\theta)$ in Eq.~\eqref{ExactMBS} are drawn by the solid line, 
and those of approximate solutions $\Psi^{\text{approx}}_{p=+,-} (\theta)$ in Eq.~\eqref{ApproxMBS} are drawn by the dashed line. 
}
\label{FigS1:MBSdensity}
\end{figure}
Fig.~\ref{FigS1:MBSdensity} shows the comparison between $\Psi^{\text{exact}}_{j=M1,M2}(\theta)$ and $\Psi^{\text{approx}}_{p=+,-} (\theta)$, 
exhibiting a quantitatively good agreement between them. This leads us to the conclusion that MBS found in Eq.~\eqref{MBSs} are exponentially 
localized in the azimuthal direction, and hence the form of the tunneling Hamiltonian of Eq. (12) in the main text is a good approximation.

\section*{B. Zero-energy Majorana bound states for non-zero chemical potential}\label{sm:nonzeromu}

We prove that Majorana bound states remain at zero energy for finite $\mu$ by using a Taylor expansion around $\mu=0$. 
We start the proof by rewriting Eq.~(1) in the main text as 
\begin{equation}
 \mathcal{H}_{\text{BdG}}(\mu) \Psi_{j}(\mu) = E_j(\mu) \Psi_{j}(\mu),\label{BdGEqnwithmu}
\end{equation}
where $E_{j=1,2}(\mu)$ are the two lowest eigenvalues which are assumed to be differentiable functions with respect to $\mu$, 
and $\Psi_{j=1,2}(\mu)$ are corresponding eigenstates. In the limit of $\mu \rightarrow 0$, $E_{j}(\mu) = 0$ and 
$\Psi_{j}(\mu)$ become two Majorana bound states, $\Psi_1(0) = \Psi_{M1}$ and $\Psi_2(0) = \Psi_{M2}$. 
Throughout this section, we use the notation $| \widetilde{\Psi} \rangle = \Xi |\Psi \rangle$ as the particle-hole transformed 
counterpart of $|\Psi \rangle$ where $\Xi = \tau_y \sigma_y \mathcal{C}$ is the particle-hole operator. 
We note that $\langle \widetilde{\Psi}'| \widetilde{\Psi} \rangle = \langle \Psi' | \Psi \rangle^*$ 
by the anti-unitary property of $\Xi$. And $| \widetilde{\Psi} \rangle = | \Psi \rangle$ if $|\Psi\rangle$ is a Majorana state.
Below, we will show that $E_{j}(\mu) = 0$ also for finite $\mu$.

The Taylor expansion of the BdG equation in terms of $\mu$ leads to
\begin{align}
 &\mathcal{H}_{\text{BdG}}(\mu) = \mathcal{H}_{\text{BdG}}(0) - \mu \tau_z, \\
 &E_j(\mu) = E_{j,0} + E_{j,1} \mu + \frac{1}{2!}E_{j,2} \mu^2 + \frac{1}{3!}E_{j,3} \mu^3 + \dots, \\
 & \Psi_{j}(\mu) = \Psi_{j,0} + \Psi_{j,1} \mu + \frac{1}{2!} \Psi_{j,2} \mu^2 + \frac{1}{3!} \Psi_{j,3} \mu^3 + \dots, 
\end{align}
where 
\begin{equation}
 E_{j,n} \equiv \frac{\partial^n E(\mu)}{\partial \mu^n} \Bigg|_{\mu=0}, \,\,\,\,\,\,\, \Psi_{j,n} \equiv \frac{\partial^n \Psi_{j}(\mu)}{\partial \mu^n} \Bigg|_{\mu=0},
\end{equation} 
and $N$ is the normalization constant. 
We note that $E_{j,0} = 0$ and $\Psi_{j,0} = \Psi_{Mj}$. By substituting these series into Eq.~\eqref{BdGEqnwithmu} and 
equating the terms in each power of $\mu$, we obtain equations for zeroth-order, first-order, 
and so on, each of which is discussed below. 

The equation for the zeroth order of $\mu$ is given by 
\begin{equation}
 \mathcal{H}_{\text{BdG}}(0) \Psi_{j,0} = E_{j,0} \Psi_{j,0} =0, \label{zerothorder}
\end{equation}
which is trivial.

Next, we consider the first-order equation given by
\begin{equation}
 \mathcal{H}_{\text{BdG}}(0) \Psi_{j,1} -\tau_z \Psi_{j,0} = E_{j,1} \Psi_{j,0}. \label{firstorder}
\end{equation}
If we multiply by $\langle \Psi_{j,0}|$, it yields 
\begin{equation}
 \langle \Psi_{j,0}| \mathcal{H}_{\text{BdG}}(0) | \Psi_{j,1} \rangle
 -\langle \Psi_{j,0}| \tau_z | \Psi_{j,0} \rangle 
 = E_{j,1} \langle \Psi_{j,0} | \Psi_{j,0} \rangle.
\end{equation}
The first term on the left side is zero by virtue of Eq.~\eqref{zerothorder}. Then the second term is real,  
and we have 
\begin{align}
 \langle \Psi_{j,0}| \tau_z | \Psi_{j,0} \rangle =& \langle \Psi_{j,0}|\tau_z  | \Psi_{j,0} \rangle^* = \langle \widetilde{\Psi}_{j,0}| \Xi \tau_z  | \Psi_{j,0} \rangle \nonumber \\
 =& -\langle \Psi_{j,0}| \tau_z | \Psi_{j,0} \rangle \nonumber\\
 =& 0,
\end{align}
resulting in $E_{j,1} =0$. Here we used anti-unitarity of $\Xi$ and anticommutation relation $\{\tau_z, \Xi\} = 0$. 
Eq.~\eqref{firstorder} with $E_{j,1} =0$ leads to 
\begin{align}
 &\mathcal{H}_{\text{BdG}}(0) \Psi_{j,1} = \tau_z \Psi_{j,0}, \label{firstorder_1}\\
 & \langle \Psi_{j',0}| \tau_z | \Psi_{j,0} \rangle =0 \,\,\,\, \forall j,j' \in \{1,2\}. \label{firstorder_2}
\end{align}
If we apply $\Xi$ to Eq.~\eqref{firstorder_1} 
\begin{align}
 \mathcal{H}_{\text{BdG}}(0) \widetilde{\Psi}_{j,1} &= \tau_z \widetilde{\Psi}_{j,0}, \nonumber\\
 & = \mathcal{H}_{\text{BdG}}(0) \Psi_{j,1}, 
\end{align}
allowing us to write 
\begin{equation}
 \widetilde{\Psi}_{j,1} = \Psi_{j,1} + \alpha_{j,1} \Psi_{1,0} + \beta_{j,1} \Psi_{2,0} \label{firstorder_3}
\end{equation}
which will be useful to prove $E_{j,2} =0$ below, where $\alpha_{j,1}$ and $\beta_{j,1}$ are constants.

We can do the similar procedure as above to get $E_{j,2} =0$. The $\mu^2$-order equation 
with $E_{j,0}= E_{j,1} =0$ is given by 
\begin{equation}
 \frac{1}{2!}\mathcal{H}_{\text{BdG}}(0) \Psi_{j,2} -\tau_z \Psi_{j,1} = \frac{1}{2!} E_{j,2} \Psi_{j,0}. \label{secondorder_1}
\end{equation}
Multiplying $\langle \Psi_{j,0}|$ to this equation gives 
\begin{equation}
 -\langle \Psi_{j,0}| \tau_z | \Psi_{j,1} \rangle 
 = E_{j,2} \langle \Psi_{j,0} | \Psi_{j,0} \rangle.
\end{equation}
Using Eqs.~\eqref{firstorder_2} and \eqref{firstorder_3}, 
\begin{align}
 \langle \Psi_{j,0}| \tau_z | \Psi_{j,1} \rangle &= \langle \Psi_{j,0}| \tau_z | \Psi_{j,1} \rangle^* 
  =-\langle \Psi_{j,0}| \tau_z | \widetilde{\Psi}_{j,1} \rangle \nonumber\\ 
  &= - \langle \Psi_{j,0}| \tau_z | \Psi_{j,1} \rangle - \alpha_{j,1} \langle \Psi_{j,0}| \tau_z | \Psi_{1,0} \rangle - \beta_{j,1} \langle \Psi_{j,0}| \tau_z | \Psi_{2,0} \rangle \nonumber\\
  &=0,
\end{align}
and we get $E_{j,2} = 0$. From Eq.~\eqref{secondorder_1}, 
\begin{align}
 &\frac{1}{2}\mathcal{H}_{\text{BdG}}(0) \Psi_{j,2} = \tau_z \Psi_{j,1} \\
 &\langle \Psi_{j',0} |\tau_z| \Psi_{j,1} \rangle = 0\,\,\,\, \forall j,j'\in \{1,2\}, \\
 & \widetilde{\Psi}_{j,2} = \Psi_{j,2} + 2 \alpha_{j,1}\Psi_{1,1} +2 \beta_{j,1}\Psi_{2,1} 
 +\alpha_{j,2} \Psi_{1,0} + \beta_{j,2} \Psi_{2,0},
\end{align}
where $\alpha_{j,2}$ and $\beta_{j,2}$ are constants.

We now verify that $E_{j,n} = 0$ for all positive integer $n$ by induction on $n$. 
It is assumed that the following conditions from the $\mu^{n}$-order equation hold: 
\begin{align}
 &E_{j,k} = 0  \,\,\,\, \text{for} \,\,\,\, 0 \leq k \leq n, \label{nthorder_1}\\
 & \mathcal{H}_{\text{BdG}}(0) \Psi_{j,n} =n \tau_z \Psi_{j,n-1} \label{nthorder_2} \\
 &\langle \Psi_{j',0} |\tau_z| \Psi_{j,k'} \rangle =0 \,\,\,\, \forall j,j'\in \{1,2\}, \,\text{and for} \,\,\,\, 0 \leq k' \leq n-1, \label{nthorder_3}\\
 & \widetilde{\Psi}_{j,n} = \Psi_{j,n} + \sum_{j'=1,2} \sum_{k'=0}^{n-1} c_{j',k'} \Psi_{j',k'}, \label{nthorder_4}
\end{align}
where $c_{j',k'}$ are constants. Since we have shown above that these conditions are fulfilled for $n \in \{1,2\}$, it is 
enough to show that if they hold for $n=n_0$, then also $n=n_0 +1$ holds. Using $E_{j,k} = 0$ for $0 \leq k \leq n_0$, 
the $\mu^{n_0 +1}$-order equation is given by 
\begin{equation}
 \frac{1}{(n_0+1)!}\mathcal{H}_{\text{BdG}}(0) \Psi_{j,n_0+1} -\frac{1}{n_0 !}\tau_z \Psi_{j,n_0} = \frac{1}{(n_0+1)!} E_{j,n_0+1} \Psi_{j,0}. \label{n0thorder}
\end{equation}
If we multiply $\langle \Psi_{j,0}|$ to this equation, we have 
\begin{equation}
 -\frac{1}{n_0 !}\langle \Psi_{j,0}| \tau_z | \Psi_{j,n_0} \rangle 
 = \frac{1}{(n_0+1)!} E_{j,n_0+1} \langle \Psi_{j,0} | \Psi_{j,0} \rangle.
\end{equation}
Using Eqs.~\eqref{nthorder_2} and \eqref{nthorder_3} with $n=n_0$, $\langle \Psi_{j,0}| \tau_z | \Psi_{j,n_0} \rangle$ in the left side can be calculated as  
\begin{align}
 \langle \Psi_{j,0}| \tau_z | \Psi_{j,n_0} \rangle &= \langle \Psi_{j,0}| \tau_z | \Psi_{j,n_0} \rangle^* 
  =-\langle \Psi_{j,0}| \tau_z | \widetilde{\Psi}_{j,n_0} \rangle \nonumber\\ 
  &= - \langle \Psi_{j,0}| \tau_z | \Psi_{j,n_0} \rangle -\sum_{j'=1,2} \sum_{k'=0}^{n_0-1} c_{j',k'} \langle \Psi_{j,0}| \tau_z | \Psi_{j',k'} \rangle \nonumber\\
  &=0,
\end{align}
and $E_{j,n_0+1} = 0$. From Eq.~\eqref{n0thorder} we get
\begin{align}
 &\mathcal{H}_{\text{BdG}}(0) \Psi_{j,n_0+1} =(n_0+1)\tau_z \Psi_{j,n_0}, \\
 & \langle \Psi_{j',0} |\tau_z| \Psi_{j,n_0} \rangle =0 \,\,\,\, \forall j,j'\in \{1,2\}, \\
 & \widetilde{\Psi}_{j,n_0+1} = \Psi_{j,n_0+1} + \sum_{j'=1,2} \sum_{k'=0}^{n_0} c'_{j',k'} \Psi_{j',k'}.
\end{align}
Therefore, if Eqs.~\eqref{nthorder_1}-\eqref{nthorder_4} hold for $n=n_0$, then they also hold for $n=n_0+1$.

\section*{C. Time-averaged tunneling current}\label{sm:current}

In this section, we provide the details of the calculation of the time-averaged tunneling current 
$\bar{I}$ in the main text. It is obtained from the average of $\langle I(t)\rangle$ over a half rotation period, 
\begin{equation}
 \bar{I} = \frac{1}{T_J} \int^{t_Q + \frac{T_J}{2}}_{t_Q -\frac{T_J}{2}} dt  \langle I(t)\rangle,
\end{equation} 
in the limit of very large integer $Q$ so that $t_Q = t_0 + Q T_J \gg t_0$. 
For a time $t \in [t_Q -T_J/2, t_Q + T_J/2]$, $\langle I(t)\rangle$ is roughly proportional 
to the tip-Majorana tunneling magnitude $\langle I(t)\rangle \propto e^{-\lambda |t-t_Q|}$. 
Thus $\langle I(t)\rangle$ is maximum when $t = t_Q$, and is exponentially small of the order of 
$e^{-\lambda T_J/2}$ for $t = t_Q \pm T_J/2$.  

We first calculate the tunneling current $\langle I(t)\rangle$ in the time interval $[t_Q -T_J/2, t_Q + T_J/2]$. 
From Eq.~(14) in the main text, we obtain
\begin{equation}
\langle I(t)\rangle = \frac{2 e}{\hbar^2} \sum_k \frac{|V_k|^2 \lambda}{\lambda^2 + (\varepsilon_k + eV)^2/\hbar^2} 
\left[-(1-n_F(\varepsilon_k)) A(t,\alpha) +  n_F(\varepsilon_k) A(t,-\alpha) \right],
\end{equation}
where 
\begin{equation}
 A(t,\alpha) = e^{2 \lambda (t-t_Q)} + e^{\lambda (t-t_Q)} \left\{ \left[\sum^{Q-1}_{n=1} e^{i n \alpha} e^{-i (\varepsilon_k+eV)(t-t_{Q-n})/\hbar} \right] 
+ \frac{\lambda + i (\varepsilon_k+eV)/\hbar}{2 \lambda} e^{i Q \alpha} e^{-i (\varepsilon_k+eV)(t-t_{0})/\hbar} + c.c. \right\} 
\end{equation}
for $t \in [t_Q -T_J/2 , t_Q]$, and 
\begin{equation}
 A(t,\alpha) = -e^{-2 \lambda (t-t_Q)} + e^{-\lambda (t-t_Q)} \left\{ \left[\sum^{Q-1}_{n=0} e^{i n \alpha} e^{-i (\varepsilon_k+eV)(t-t_{Q-n})/\hbar} \right] 
+ \frac{\lambda + i (\varepsilon_k+eV)/\hbar}{2 \lambda} e^{i Q \alpha} e^{-i (\varepsilon_k+eV)(t-t_{0})/\hbar} + c.c. \right\} 
\end{equation}
for $t \in [t_Q, t_Q +T_J/2]$. $A(t,-\alpha)$ is obtained by replacing $\alpha$ by $-\alpha$ from $A(t,\alpha)$. 
In this calculation, we neglected the terms of the order of $e^{-\lambda T_J/2}$ that give an exponentially 
small contribution to $A(t,\alpha)$ in our limit of the short duration of the tunneling $\lambda^{-1} \ll T_J$. 
Then $\bar{I}$ can be expressed as 
\begin{equation}\label{AverageI}
\bar{I} = \frac{2 e}{\hbar^2} \sum_k \frac{|V_k|^2 \lambda}{\lambda^2 + (\varepsilon_k + eV)^2/\hbar^2} \left\{ 
-(1-n_F(\varepsilon_k)) \left[ \frac{1}{T_J} \int^{t_Q + \frac{T_J}{2}}_{t_Q -\frac{T_J}{2}} dt A(t,\alpha) \right] 
+  n_F(\varepsilon_k) \left[  \frac{1}{T_J} \int^{t_Q + \frac{T_J}{2}}_{t_Q -\frac{T_J}{2}} dt A(t,- \alpha) \right] \right\},
\end{equation}

To calculate $\bar{I}$ we have to calculate the time averages of $A(t,\alpha)$ and $A(t,-\alpha)$.  
It is given by 
\begin{align}
 \frac{1}{T_J} \int^{t_Q + \frac{T_J}{2}}_{t_Q -\frac{T_J}{2}} dt A(t,\alpha) = \frac{1}{T_J} 
 \Bigg\{&\frac{2 \lambda}{\lambda^2 + (\varepsilon_k + eV)^2/\hbar^2} \left[\sum^{Q}_{n= -Q} e^{i n \alpha} e^{-i (\varepsilon_k+eV)n T_J/\hbar} \right] \nonumber\\
 &+ \left[ \frac{-\lambda + i (\varepsilon_k + eV)/\hbar}{\lambda^2 + (\varepsilon_k + eV)^2/\hbar^2} e^{i \alpha Q} e^{-i (\varepsilon_k+eV)Q T_J/\hbar} +c.c. \right] \Bigg\}.
\end{align}
In the limit that $Q$ goes to infinity, the summation term over $n$ on the right side can be written as 
\begin{align}
 \lim_{Q\rightarrow \infty} \sum^{Q}_{n= -Q} e^{i n \alpha} e^{-i (\varepsilon_k+eV)n T_J/\hbar} =& 2 \pi \sum^{\infty}_{l=-\infty} 
 \delta \left[ \alpha - \frac{(\varepsilon_k + eV) T_J}{\hbar} - 2 \pi l \right] \nonumber\\
 =& \frac{2 \pi \hbar}{T_J} \sum^{\infty}_{l=-\infty} \delta \left[- \varepsilon_k - eV + \frac{(\alpha- 2\pi l) \hbar}{T_J} \right],
\end{align}
where $l = 0, \pm1, \pm2,...$, while the remaining terms that possess $e^{\pm i \varepsilon_k Q T_J/\hbar}$ factors can be neglected because of their rapid oscillation 
as a function of $k$ for large $Q$ 
\begin{equation} 
 \sum_{k} e^{\pm i \varepsilon_k Q T_J/\hbar} \sim 0.
\end{equation}
Thus the time averages of $A(t,\alpha)$ and $A(t,-\alpha)$ (obtained by replacing $\alpha$ by $-\alpha$ in $A(t,\alpha)$) have the forms 
\begin{align}
 \frac{1}{T_J} \int^{t_Q + \frac{T_J}{2}}_{t_Q -\frac{T_J}{2}} dt A(t,\alpha) &= 
 \frac{2 \lambda}{\lambda^2 + (\varepsilon_k + eV)^2/\hbar^2} \frac{2 \pi \hbar}{T^2_J} \sum^{\infty}_{l=-\infty} \delta \left[- \varepsilon_k - eV + \frac{(\alpha- 2\pi l) \hbar}{T_J} \right], \nonumber\\ \nonumber\\
 \frac{1}{T_J} \int^{t_Q + \frac{T_J}{2}}_{t_Q -\frac{T_J}{2}} dt A(t,-\alpha) &= 
 \frac{2 \lambda}{\lambda^2 + (\varepsilon_k + eV)^2/\hbar^2} \frac{2 \pi \hbar}{T^2_J} \sum^{\infty}_{l'=-\infty} \delta \left[- \varepsilon_k - eV - \frac{(\alpha + 2\pi l') \hbar}{T_J} \right], \nonumber\\
 &=\frac{2 \lambda}{\lambda^2 + (\varepsilon_k + eV)^2/\hbar^2} \frac{2 \pi \hbar}{T^2_J} \sum^{\infty}_{l'=-\infty} \delta \left[- \varepsilon_k - eV - \frac{(\alpha - 2\pi l') \hbar}{T_J} \right].
\end{align}
In the last line of the second equation, we changed $l'$ by $-l'$ for convenience. 
If we put these into Eq.~\eqref{AverageI}, it follows that 
\begin{align}
 \bar{I} = \frac{e}{\hbar^2} \sum_{k} |V_k|^2\frac{2 \pi \hbar}{T^2_{J}} \left[ \frac{2 \lambda}{\lambda^2 + (\varepsilon_k + eV)^2/\hbar^2} \right]^2 
 \Bigg\{& -(1-n_F(\varepsilon_k)) \sum^{\infty}_{l=-\infty} \delta \left[- \varepsilon_k - eV + \frac{(\alpha- 2\pi l) \hbar}{T_J} \right] \nonumber\\
 & + n_F(\varepsilon_k) \sum^{\infty}_{l'=-\infty} \delta \left[- \varepsilon_k - eV - \frac{(\alpha - 2\pi l') \hbar}{T_J} \right]  \Bigg\}.
\end{align}
By replacing the discrete sum over $k$ by an integral with the density of states of the metallic tip $\rho(\varepsilon_k)$
\begin{equation}
 \sum_k \rightarrow \int d \varepsilon_k \rho(\varepsilon_k).
\end{equation}
Assuming a wide-band limit with an energy independent tunneling rate $\Gamma = 2 \pi \rho(\varepsilon_k) |V_k|^2$, we obtain
\begin{align}
\bar{I} =& \frac{e}{h}\sum_{l} 2\pi \Gamma \left[ \frac{2 \lambda T_J}{\lambda^2 T^2_J + (\alpha - 2\pi l)^2}\right]^2 
\left[-n_F(eV -(\alpha - 2 \pi l) \hbar /T_J) + n_F(-eV -(\alpha - 2 \pi l) \hbar /T_J) \right] \nonumber\\
=& \frac{e}{h} \sum_{l} Z_l(\alpha) \left[-n_F(V^{+}_l (\alpha)) + n_F(V^{-}_l (\alpha)) \right],
\end{align}
where $Z_l(\alpha)$ and $V^{\pm}_l (\alpha)$ are defined in the main text. 
It is now straightforward to evaluate the time-averaged differential conductance $d \bar{I}/dV$. Since only the Fermi-Dirac distributions 
have the voltage dependence, it is enough to compute 
\begin{align}
 \frac{d n_F(V^{\pm}_l (\alpha))}{dV} =& \frac{d}{dV} \frac{1}{1+ \text{exp} \left[V^{\pm}_l (\alpha)/(k_B T) \right]} \nonumber\\
 =& \mp \frac{e}{4 k_B T} \text{sech}^2 \left[ V^{\pm}_l (\alpha)/(2 k_B T) \right].
\end{align}
Therefore, we get 
\begin{equation}
 \frac{d \bar{I}}{dV} =\frac{e^2}{h} \sum_{l} \frac{Z_l(\alpha)}{4 k_B T} \left\{\text{sech}^2 \left[ \frac{V^{+}_l(\alpha)}{2 k_B T} \right] 
 +\text{sech}^2 \left[ \frac{V^{-}_l(\alpha)}{2 k_B T} \right] \right\}.
\end{equation}

\section*{D. Floquet description of the two rotating Majorana bound states}

In the previous section, we showed that the time-averaged differential conductance has peaks at 
$eV = \pm (\alpha - 2 \pi l) \hbar/T_J$. Here, we provide an alternative derivation of the result 
by considering an effective Floquet Hamiltonian for the two rotating Majorana bound states in our Corbino-geometry Josephson junction. 
The Floquet description is valid as the time-dependent BdG Hamiltonian in Eq.(7) in the main text is periodic in time with period $T_J$: 
$H_{\text{BdG}}[\Delta\phi(t')] = H_{\text{BdG}}[\Delta\phi(t' + T_J)]$ where $\Delta\phi$ is the phase difference 
across the junction. The time evolution operator of the Hamiltonian from $t'$ to $t'+T_J$ in our low-energy subspace is $U_M$ 
in Eq. (17) in the main text, and the effective Floquet Hamiltonian $H_F$~\cite{Cayssol2013_SM} is obtained by using 
\begin{equation}
 e^{-i H_F T_J / \hbar} = U_M, 
\end{equation}
which yields 
\begin{equation}
 H_F = i \frac{\varepsilon}{2} \gamma_2(t_0) \gamma_1(t_0) + 2 \pi l \frac{\hbar}{T_J},  
\end{equation}
where $\varepsilon = \alpha \hbar / T_J$ and integer $l$. The first term is reminiscent to the Hamiltonian 
for two coupled static MBS with 
hybridisation $\varepsilon$ (which shows conductance peaks at $eV = \pm \varepsilon$), and 
the second term shifts energy levels. As a result, tunneling into an effective system described by $H_F$ gives 
conductance peaks at $eV = \pm \varepsilon + 2 \pi l \hbar / T_J$ which is producing our features above.

\section*{E. Weak disorder effect}\label{sm:disorder} 

In this section, we show that two Majorana bound states in a Corbino-geometry Josephson junction are robust 
against weak disorder on the surface of a topological insulator if we stay in the subspace spanned by the two Majorana 
bound states, i.e., they remain at zero energy.
It can be shown by using a Taylor expansion in the same way as in Section B. 
The surface of a topological insulator with a disorder potential is described by 
\begin{align}
 H = v_{\text{F}} \vec{\sigma} \cdot \vec{p} + V(\vec{r}),
\end{align}
where $V(\vec{r})$ is the disorder potential. Then the BdG equation is given by 
\begin{align}
 \left[\left(
 \begin{array}{cc}
  v_{\text{F}} \vec{\sigma} \cdot \vec{p} & \,\, \Delta(r, \theta) \mathbb{I} \\
  \Delta^{*}(r, \theta) \mathbb{I} & \,\, -v_{\text{F}} \vec{\sigma} \cdot \vec{p}
 \end{array}
\right) + 
\left(
 \begin{array}{cc}
  \lambda V(\vec{r}) & 0 \\
  0 & - \lambda V(\vec{r})
 \end{array} 
\right) \right]\Psi'_j(r, \theta) = E'_j \Psi'_j(r, \theta). 
\end{align}
Here a perturbation parameter $\lambda$ is introduced to keep track of the order of the disorder potential in a Taylor expansion, and  
$\Psi'_{j=1,2}(r, \theta)$ and $E'_j$ are eigenfunctions and eigenvalues, respectively, which are assumed to be differentiable 
functions with respect to $\lambda$.
Like in the case of the calculation in Section B, we do a Taylor expansion for $\Psi'_{j=1,2}(r, \theta)$ and $E'_j$ in $\lambda$, 
\begin{align}
 &E'_j = E'_{j,0} + \lambda E'_{j,1} + \frac{\lambda^2}{2!}E'_{j,2} + \frac{\lambda^3}{3!}E'_{j,3} + \dots, \\
 & \Psi'_{j} = \Psi'_{j,0} + \lambda \Psi'_{j,1} + \frac{\lambda^2}{2!} \Psi'_{j,2} + \frac{\lambda^2}{3!} \Psi'_{j,3} + \dots. 
\end{align}
Now we substitute these expansions into the BdG equation and equate terms of the same order in $\lambda$. 
All calculation details are the same as in Section B, except that 
$\mu$ and $-\tau_z$ in Section B are replaced by $\lambda$ and $V(\vec{r}) \tau_z$ where $\tau_z$ is the Pauli matrix in particle-hole space, 
and we get 
\begin{align}
 E'_{j,k} = 0,
\end{align}
for all $j \in \{1,2 \}$ and nonnegative integer $k$. Therefore we conclude that two Majorana bound states are robust against a weak disorder potential, 
and hence the results shown in Fig. 3 in the main text can be detected even in the presence of weak disorder.


\begin{thebibliography}{50}

\bibitem{Kitaev2001} A. Y. Kitaev, Phys. Usp. \textbf{44}, 131 (2001). 

\bibitem{Hasan2010} M. Z. Hasan and C. L. Kane, Rev. Mod. Phys. \textbf{82}, 3045 (2010). 

\bibitem{Alicea2012} J. Alicea, Rep. Prog. Phys. \textbf{75}, 076501 (2012).

\bibitem{Beenakker2013} C. W. J. Beenakker, Annu. Rev. Condens. Matter Phys. \textbf{4}, 113 (2013). 

\bibitem{Nayak2008} C. Nayak, S. H. Simon, A. Stern, M. Freedman, and S. Das Sarma, 
Rev. Mod. Phys. \textbf{80}, 1083 (2008).

\bibitem{Moore1991} G. Moore and N. Read, Nucl. Phys. \textbf{B360}, 362 (1991).

\bibitem{Ivanov2001} D. A. Ivanov, Phys. Rev. Lett \textbf{86}, 268 (2001).

\bibitem{Fu2008} L. Fu and C. L. Kane, Phys. Rev. Lett \textbf{100}, 096407 (2008). 


\bibitem{Sau2010} J. D. Sau, R. M. Lutchyn, S. Tewari, and S. Das Sarma, Phys. Rev. Lett \textbf{104}, 040502 (2010).


\bibitem{Oreg2010} Y. Oreg, G. Refael, and F. von Oppen, Phys. Rev. Lett. \textbf{105}, 177002 (2010). 


\bibitem{Choy2011} T.-P. Choy, J. M. Edge, A. R. Akhmerov, and C. W. J. Beenakker, Phys. Rev. B \textbf{84}, 195442 (2011). 

\bibitem{Nadj-Perge2013} S. Nadj-Perge, I. K. Drozdov, B. A. Bernevig, and A. Yazdani, Phys. Rev. B \textbf{88}, 020407(R) (2013). 

\bibitem{Nadj-Perge2014} S. Nadj-Perge, I. K. Drozdov, J. Li, H. Chen, S. Jeon, J. Seo, A. H. MacDonald, B. A. Bernevig, and A. Yazdani,
Science \textbf{346}, 602 (2014).

\bibitem{Law2009} K. T. Law, P. A. Lee, and T. K. Ng, Phys. Rev. Lett. \textbf{103}, 237001 (2009). 

\bibitem{Flensberg2010} K. Flensberg, Phys. Rev. B \textbf{82}, 180516(R) (2010). 

\bibitem{Mourik2012} V. Mourik , K. Zuo, S. M. Frolov, S. R. Plissard, E. P. A. M. Bakkers, and L. P. Kouwenhoven, 
Science \textbf{336}, 1003 (2012). 

\bibitem{Das2012} A. Das, Y. Ronen, Y. Most, Y. Oreg, M. Heiblum, and H. Shtrikman, Nat. Phys. \textbf{8}, 887 (2012). 

\bibitem{Deng2012} M.T. Deng, C.L. Yu, G.Y. Huang, M. Larsson, P. Caroff, and H.Q. Xu, Nano Lett. {\bf 12}, 6414 (2012). 

\bibitem{Lee2014} E. J. H. Lee, X. Jiang, M. Houzet, R. Aguado, C. M. Lieber, and S. De Franceschi, Nat. Nanotechnol. \textbf{9}, 79 (2014).

\bibitem{Fu2009PRL} L. Fu and C. L. Kane, Phys. Rev. Lett. \textbf{102}, 216403 (2009)

\bibitem{Akhmerov2009} A. R. Akhmerov, J. Nilsson, and C. W. J. Beenakker, Phys. Rev. Lett. \textbf{102}, 216404 (2009). 

\bibitem{Fu2009PRB} L. Fu and C. L. Kane, Phys. Rev. B \textbf{79}, 161408(R) (2009).

\bibitem{Ioselevich2011} P. A. Ioselevich and M. V. Feigel'man, Phys. Rev. Lett. \textbf{106}, 077003 (2011). 

\bibitem{Jiang2011} L. Jiang, D. Pekker, J. Alicea, G. Refael, Y. Oreg, and F. von Oppen, 
Phys. Rev. Lett. \textbf{107}, 236401 (2011). 

\bibitem{Rokhinson2012} L. P. Rokhinson, X. Liu, and J. K. Furdyna, 
Nat. Phys. \textbf{8}, 795 (2012).

\bibitem{Sacepe2011} B. Sac\'{e}p\'{e}, J. B. Oostinga, J. Li, A. Ubaldini, N. J. G. Couto, E. Giannini, and A. F. Morpurgo, 
Nat. Commun. \textbf{2}, 575 (2011). 

\bibitem{Williams2012} J. R. Williams, A. J. Bestwick, P. Gallagher, S. S. Hong, Y. Cui, A. S. Bleich, J. G. Analytis, I. R. Fisher, and D. Goldhaber-Gordon, 
Phys. Rev. Lett. \textbf{109}, 056803 (2012). 

\bibitem{Veldhorst2012} M. Veldhorst, M. Snelder, M. Hoek, T. Gang, V. K. Guduru, X. L. Wang, U. Zeitler, W. G. van der Wiel, A. A. Golubov, H. Hilgenkamp, and A. Brinkman, 
Nat. Mater. \textbf{11}, 417 (2012). 

\bibitem{Alicea2011} J. Alicea, Y. Oreg, G. Refael, F. von Oppen, and M. P. A. Fisher, Nat. Phys. \textbf{7}, 412 (2011). 

\bibitem{Heck2012} B. van Heck, A. R. Akhmerov, F. Hassler, M. Burrello, and C. W. J. Beeankker, 
New J. Phys. \textbf{14}, 035019 (2012).

\bibitem{Li2014}J. Li, T. Neupert, B. A. Bernevig, and A. Yazdani, 
arXiv:1404.4058. 

\bibitem{Karzig2015}T. Karzig, F. Pientka, G. Refael, and F. von Oppen, Phys. Rev. B \textbf{91}, 201102(R) (2015).

\bibitem{Grosfeld2011} E. Grosfeld and A. Stern, Proc. Natl. Acad. Sci. U.S.A. \textbf{108}, 11810 (2011).

\bibitem{Mechanism} Although Majorana exchange in a system of two MBS $\gamma_1$ and $\gamma_2$ 
cannot change the occupation of the non-local fermion $f=(\gamma_1+i\gamma_2)/2$ in Ref.~\cite{Ivanov2001}, 
our tunneling setup shown in Fig.~\ref{Fig1:setup} detects the exchange phase of two MBS via an interference effect 
between different orders of exchange (or braiding) cycles.

\bibitem{Clem2010} J. R. Clem, Phys. Rev. B \textbf{82}, 174515 (2010). 

\bibitem{Akzyanov2014} R. S. Akzyanov, A. V. Rozhkov, A. L. Rakhmanov, and F. Nori, 
Phys. Rev. B \textbf{89}, 085409 (2014).

\bibitem{Rakhmanov2011} A. L. Rakhmanov, A. V. Rozhkov, and F. Nori, 
Phys. Rev. B \textbf{84}, 075141 (2011).

\bibitem{Supp} See the Supplemental Material, which includes Refs.~\cite{Fu2008, Cayssol2013}, for further details on the derivation 
of MBS, the calculation of $\bar{I}$, and the effect of weak disorder. 

\bibitem{Cayssol2013}
J. Cayssol, B. D{\'o}ra, F. Simon, and R. Moessner, Phys. Status Solidi RRL \textbf{7}, 101 (2013).

\bibitem{Potter2013} A. C. Potter and L. Fu, Phys. Rev. B \textbf{88}, 121109(R) (2013).


\bibitem{periodicity} Note that $\mathcal{H}_{\text{BdG}}(\phi_1(t),\phi_2(t))$ is periodic in time 
with periodicity $T_J$ if either $\phi_1$ or $\phi_2$ is changed in time.

\bibitem{switchon}
Different switching-on times give the same current as long as they are far away from $t$ (by many exchange periods $T_J$).

\bibitem{TipGreenftn} The tip electron Green's functions are given by 
$i G_{k}(t,t') = \langle \hat{c}_{k}(t) \hat{c}^{\dagger}_{k}(t')\rangle 
= e^{-i (\varepsilon_k + e V) (t-t')/\hbar} [1-n_F(\varepsilon_k)]$, and  
$i \bar{G}_{k}(t,t') = \langle \hat{c}^{\dagger}_{k}(t) \hat{c}_{k}(t')\rangle
= e^{i (\varepsilon_k + e V) (t-t')/\hbar} n_F(\varepsilon_k)$ where 
$n_F(\varepsilon_k)= 1/[1+ e^{\varepsilon_k/(k_B T)}]$ 
is the Fermi-Dirac distribution at $t=t_0$, and tip voltage $V$



\bibitem{Lin2013}
S.-Z. Lin, O. Ayala-Valenzuela, R. D. McDonald, L. N. Bulaevskii, T. G. Holesinger, F. Ronning, N. R. Weisse-Bernstein, 
T. L. Williamson, A. H. Mueller, M. A. Hoffbauer, M. W. Rabin, and M. J. Graf, Phys. Rev. B \textbf{87}, 184507 (2013).

\bibitem{Rainis2012} D. Rainis and D. Loss, Phys. Rev. B \textbf{85}, 174533 (2012).


\end{thebibliography}

\begin{thebibliography}{10}


\bibitem{Fu2008_SM} L. Fu and C. L. Kane, Phys. Rev. Lett \textbf{100}, 096407 (2008). 

\bibitem{Cayssol2013_SM}
J. Cayssol, B. D{\'o}ra, F. Simon, and R. Moessner, Phys. Status Solidi RRL \textbf{7}, 101 (2013).

\end{thebibliography}
\end{document}